\documentclass[12pt,a4paper]{article}
\usepackage{epsfig}
\usepackage{a4}
\usepackage{graphics}

\newcommand{\bmath}[1]{\mbox{\boldmath ${#1}$}}

\newcommand{\half}{\mbox{${\textstyle \frac{1}{2}}$}}           
\newcommand{\fourth}{\mbox{${\textstyle \frac{1}{4}}$}}         

\newcommand{\dd}{\textrm{d}}
\begin{document}
\baselineskip 4ex
\begin{titlepage}
\vspace*{-2.5cm}
\hfill\textbf{TSL/ISV-2001-0246}
\vspace{5mm}

\begin{center}
\vspace*{10mm}
{\Large{\bf The production of $\bmath{\eta}$-mesons in nucleon-nucleon}}
\\[2ex]
{\Large {\bf  collisions near threshold}}
\\[6ex]
{\large G\"{o}ran F\"{a}ldt}\footnote{Electronic address: faldt@tsl.uu.se}
\\[1ex]
{\normalsize  Division of Nuclear Physics,  Box 535, 751 21 Uppsala, Sweden}
\\[3ex]
{\large Colin Wilkin}\footnote{Electronic address: cw@hep.ucl.ac.uk}\\[1ex]
{\normalsize Department of Physics \& Astronomy, UCL, London WC1E 6BT, UK}
\\[3ex]
\today\\[3ex]
\end{center}

\begin{abstract}
Data on the total cross sections for the $pp\to pp\eta$,
$pn\to pn\eta$, and $pn\to d\eta$ reactions and the $pp\to pp\eta$
differential cross section near threshold are analysed in a
one-meson-exchange model. After including initial and final-state
nucleon-nucleon distortion, the magnitude and most of the energy
dependence are well reproduced. It is found that the contribution of
$\rho$-exchange is larger than that of $\pi$-exchange. With 
destructive $\rho/\pi$
interference in the $pp$ case, the model explains quantitatively the
$pp\to pp\eta/pn\to pn\eta$ cross section ratio and the slope of the
$pp\to pp\eta$ differential cross section. Such an agreement would be
destroyed by any significant $\eta$-exchange term. The residual energy
dependence may be associated with $\eta$-nucleon rescattering that has
not been taken into account. The $pn\to pn\eta$/$pn \to d\eta$ ratio
depends weakly upon the nature of the particle exchanges, being determined
primarily by the nucleon-nucleon final state interactions. The proton
analysing power is predicted to remain small in the low energy region.
\end{abstract}
\vspace{5mm}
\noindent

\vfill
\baselineskip 2ex
\noindent
{\small Corresponding author:\\
Colin Wilkin,\\
Physics \& Astronomy Dept.,\\
UCL, Gower St.,\\
London WC1E 6BT, U.K.}
\end{titlepage}

\baselineskip 4ex
\setlength{\unitlength}{1mm}
\section{Introduction}

The experimental database on $\eta$ production in nucleon-nucleon
scattering has expanded significantly in recent years. The excitation
function of the $pp\to pp\eta$ total cross section shows such a rapid
rise with excess energy $Q$~\cite{Pinot,Calen1,Hibou,Cosy11}, that the
major error is often associated with the determination of $Q$ rather
than of the cross section itself. Equally striking is the large cross
section found for quasi-free $\eta$ production on a deuterium
target~\cite{Pinot}; under conditions of well-controlled kinematics the
$pn\to pn\eta/pp\to pp\eta$ ratio is found to be over 6
near threshold~\cite{Calen5,Stina}. The CELSIUS group has also measured the
$pn \to d\eta$ total cross section at well-defined c.m.\ 
energies~\cite{Stina,Calen2} and found the first evidence
of $N^*(1535)$ dominance from the energy dependence. More tantalisingly,
they showed that at very small $Q$ there appears to be a threshold
enhancement~\cite{Calen3} which, though not as spectacular as that found
in the analysis~\cite{PFW} of early Saclay data, nevertheless indicates
a strong $\eta$-deuteron scattering length.

The only measurement of the angular distribution of $\eta$-mesons near
threshold suggests that $d$-waves cannot be neglected in the $\eta$-$(NN)$
system for $Q\approx 30$~MeV~\cite{Calen4}. It is here important to note
that the differential cross section has the opposite curvature to that of
$\pi^- p\to \eta\,n$ and we will argue that this helps to identify the
reaction mechanism. In view of these new data, as well as the experimental
advances in $\pi^-p\to \eta\,n$ and especially $\gamma p\to\eta\,p$, we
believe that it is useful to revisit the phenomenological analysis of
these reactions.

In the usual theoretical approach, the $N^*(1535)$ (or other) isobar 
is excited in nucleon-nucleon collisions through the exchange of a 
single meson $X$~\cite{GW,Laget,Vetter,Moalem,Alfred1,PGR,Grishina}. 
The $N^*(1535)$ has a large branching ratio into $\eta\,N$ so that,
after its decay, one is left with an $\eta$ plus two nucleons in the
final state. Though the philosophies are generally rather similar, the
theoretical calculations differ in their details as to which exchanges
are relevant, the structure of their coupling, and the importance of
the associated form factors. Different techniques have been used in
order to include the effects of the initial nucleon-nucleon distortion
and the final-state interactions. The neutron-proton final-state
interaction can also produce a deuteron in the exit channel and, as we
shall see, the ratio of the $pn\to d\eta$ and $pn\to np\eta$ cross 
sections is determined primarily by low-energy neutron-proton dynamics 
with a weaker dependence upon the production mechanism~\cite{FW3}.

It is generally agreed that in the near-threshold region, $Q\leq 40$~MeV, 
the energy variation of the total cross section for the $NN\eta$ 
channel is fixed mainly by the $Q^2$ factor coming from phase space, modified
by the nucleon-nucleon final-state interaction. It is known, however, that
$\eta$ rescattering does produce threshold enhancements in the
$pd\to\,^3$He$\,\eta$~\cite{Mayer} and
$dd\to\,^4$He$\,\eta$~\cite{Willis} reactions and it is likely that
the residual energy dependence in the $pp\to pp\eta$ excitation function
could be due to such an effect. The energy dependence coming from the input
single-nucleon production amplitudes is rather modest over a small range
in $Q$.

The relativistic Born amplitudes for single-meson exchange are first 
evaluated at threshold by neglecting all distortions. To include the
effects of the final-state interactions it is, however, necessary to
construct a potential from the amplitudes and this is most easily done
in configuration space. The energy and angular dependence coming from
$\pi^-p\to \eta\,n$ and the other elementary amplitudes can then be
introduced in a perturbation approach.

The kinematics of the processes are outlined in Section~2. As
exchanged particles, we consider the $\pi$, $\eta$, $\rho$, and
$\omega$. The forms of their coupling to the nucleon are given in
Section~3, together with the coupling constants used in this work,
though it must be stressed that the uncertainty in the $\eta NN$ value
is very large. Values for the $XN \to \eta\,N$ amplitudes are also 
discussed in this section. There are, of course, measurements of the
pion-induced cross section and the $\eta$-nucleon elastic cross section
can be deduced from this inside a unitary model. To obtain the necessary
vector meson information, we interpret photoproduction
data within the framework of the vector meson dominance approach.

The formulae for the bare meson exchange $NN\to NN\eta$ and
$pn\to d\eta$ amplitudes are derived in Section~4; the energy and 
angular dependence coming from the $XN \to \eta\,N$ amplitudes are there
introduced.

Section~5 is devoted to the evaluation of the final state
interactions in the $pp$ and $np$ systems. It is found that the
enhancement factor varies very fast with excitation energy, and also
depends upon isospin in the $NN$ channel as well as on the mass of 
the exchanged meson. In contrast, the distortion of the initial 
nucleon-nucleon wave is slowly varying and we try to include its
effects simply by using the imaginary part of the relevant phase shift. 
The $\eta$-nucleon final-state interaction is not taken into account, 
since it would require a consistent three-body treatment in order to 
include this simultaneously with the nucleon-nucleon interaction. There 
is, moreover, as yet no credible $\eta$-nucleon potential from which 
to calculate the wave function at short distances. The two-body 
$pn\to d\eta$ amplitude is discussed in Section~6.

As shown in Section~7, $\rho$-exchange is more important than
$\pi$-exchange, with the $\omega$ giving only minor modifications. Taking
the $\rho/\pi$ interference to be destructive in the proton-proton
case, these three exchanges reproduce well the measured $6.5\!:\!1$ $pn\!:\!pp$
ratio. Since pure $\eta$ exchange would lead to a ratio of below 1,
the results yield an upper bound on the poorly determined $\eta$-nucleon
coupling constant. With standard parameters, the model reproduces well the
magnitude and most of the energy dependence observed in the total cross.
There is an indication that the model underpredicts the $pp\to pp\eta$
data at low $Q$ and the same is true for the two-body $pn\to d\eta$
results. This is probably due to the neglect of $\eta$ rescattering.
On the other hand, the reasonable agreement found for the 
$\sigma(pn\to pn\eta)/\sigma(pn\to d\eta)$ cross section ratio is
mainly a reflection of the $np$ final state interaction.

The dominance of the $\rho$ and the destructive effect of the $\pi$
explains also the shape of the $\eta$ angular distribution in
$pp\to pp\eta$,
that is larger at $90^{\circ}$ than in the forward direction~\cite{Calen4}.
There are still insufficient input data to give definitive predictions for
the proton analysing power, but the indications are that this should be
rather small in the near-threshold region. Our conclusions are given in
Section~8.
\newpage
\section{Kinematics}

We wish to describe the $pp\to pp\eta$, $pn\to np\eta$ and 
$pn\to d\eta$ reactions in the near-threshold domain. The energies of the 
incident nucleons in the c.m.\ system are denoted by $E$ and their 
momenta $\pm\bmath{p}$. At threshold they are related to the nucleon
and $\eta$ masses, $M$ and $\mu$, by
\begin{eqnarray}
\nonumber
 E &=& M+ \half \mu  \\
 p &=& \sqrt{\mu M+ \fourth \mu^2}\:.
\label{2_1}
\end{eqnarray} 
The excess energy $Q$ is fixed by the total c.m.\ energy as 
$W=2E = 2M +m_{\eta}+Q$, where we shall neglect the neutron-proton mass
difference except in the determination of $Q$ from the beam energy. 
In the case of the two-body $pn\to d\eta$ reaction, the $2M$ is replaced
by the deuteron mass, $M_d$. Reduced masses in the final state 
are defined by
\begin{equation}
\mu_{\eta}= \frac{\mu}{1+\mu/2M} \hspace{1cm}
\mu_{12}=M/2\:.
\label{2_2}
\end{equation}
As an alternative to $Q$, experimentalists often quote the value of the
maximum $\eta$ momentum, $p_{\eta}$, in units of the $\eta$ mass;
\begin{equation}
\label{2_3}
\eta=\frac{p_{\eta}^{\textrm{\scriptsize max}}}{\mu}=
\sqrt{\frac{2\mu_{\eta}Q}{\mu^2}}
\:\cdot
\end{equation}

If the matrix element $\displaystyle\cal{M}$ were constant at its
threshold value, the only energy dependence in the
unpolarised total cross section would come from the $Q^2$ factor
arising from phase space. Thus, for the reaction $pp\rightarrow pp\eta$,
\begin{equation}
\sigma(pp\rightarrow pp\eta)= \frac{1}{64\pi^2 pW}\:
  \frac{(\mu_{\eta}\,\mu_{12})^{3/2}}{8M^2\mu}\: Q^2 \  
 \left[\fourth \sum | {\cal{M}}| ^{2} \right]\:, 
\label{2_4}
\end{equation}
where the summation is over the initial and final spin projections.
This expression incorporates a factor of one-half coming from 
the identity of the final-state protons. Such a factor is absent from the 
corresponding $pn\rightarrow pn\eta$ formula.

Finally, we also need the cross section for $pn\rightarrow d\eta$,
which is      
\begin{equation}
\sigma(pn\rightarrow d\eta)= \frac{1}{16\pi W^2}\frac{p_{\eta}}{p}\: 
 \left[\fourth \sum | {\cal{M}}| ^{2} \right] \:.
\label{2_5}
\end{equation}

As we shall see later, the strong nucleon-nucleon final state
interaction will lead to a major modification in the $Q^2$ behaviour
such that it will no longer be permissible to factorise the dynamics
from the phase space, as has been done in Eq.~(\ref{2_4}). 
\newpage
\section{Input Amplitudes and Vertices}
The basic inputs required for the evaluation of our Feynman diagrams 
are vertex functions and $\eta$-production amplitudes
$\cal{M}$, which are the matrix elements of 
$i{\cal L}_{\rm int}=-i{\cal H}_{\rm int}$. We list 
below their general forms as well as the corresponding  
approximations used in the evaluation of the near-threshold amplitudes. 
\subsection{Pseudoscalar meson-nucleon vertex}
The standard $\pi NN$ coupling in terms of four-component spinors $u(p)$ is
\begin{equation}
{\cal M}_{\pi NN}= -\frac{f_{\pi}}{m_{\pi}}\ 
\bar{u}(\bmath{p}')\gamma_5(\not\!p\,'-\not\!p)\,
\bmath{\tau}\cdot \bmath{\phi}_{\pi}\,u(\bmath{p})\:,
\label{3_1_1}
\end{equation}
where we take the pion-nucleon coupling
\begin{equation} 
G_{\pi}=\frac{2Mf_{\pi}}{m_{\pi}}
\label{3_1_2}
\end{equation}
to have the numerical value $G_{\pi}^2/4\pi=13.6$~\cite{Machleidt}. 
Since we are specialising to near-threshold reactions, it is sufficient
in our applications to put $\bmath{p}'=\bmath{0}$ and $E'=M$ in the final 
state. 

It is important to take into account the off-shell extrapolation in the 
vertex function and for this we introduce a monopole form factor
\begin{equation}
F_{\pi}(q^2)=\frac{\Lambda_{\pi}^2-m_{\pi}^2}{ \Lambda_{\pi}^2 - q^2}\:,
\label{3_1_3}
\end{equation}
where $q^2$ is the square of the pion four-momentum. We take 
the value $\Lambda_{\pi}=1.72$~GeV/c, as recommended in \cite{Machleidt}.

The size of the $\eta$-nucleon coupling constant is extremely uncertain,
with values of $G_{\eta}^{\,2}/4\pi$ between 0 and 7 being quoted in the
literature~(see \textit{e.g.}~\cite{Zhu}). A value of 0.4 has been
deduced from fits to photoproduction~\cite{Tiator2}. However, even with 
a value as small as that, it is claimed that $\eta$-exchange would 
strongly influence $\eta$ production in
nucleon-nucleon collisions~\cite{PGR}. Since pure $\eta$-exchange gives
a completely wrong prediction for the ratio of the relative production
in the $pp$ and $pn$ cases, these data will provide evidence for some upper
limit for $G_{\eta}^{\,2}/4\pi$ within the framework of the meson-exchange
model for $\eta$ production.

The range parameter in the $\eta$ form factor will be taken, as in the
Bonn meson-exchange potential~\cite{Bonn}, to have the value
$\Lambda_{\eta}=1.5$~GeV.
\subsection{Vector-meson-nucleon vertex}
The generic form of the couplings of vector mesons to the nucleon is
(see \textit{e.g.}~\cite{Nimai})
\begin{eqnarray}
\nonumber
{\cal M}_{V NN}&=&- ig_V\  
\bar{u}(\bmath{p}')\left[\not\!{\epsilon} + \frac{\kappa_V}{4M}
  (\not\!\epsilon \not\!k
-\not\!k\not\!\epsilon)\right]\,u(\bmath{p})\:,\\
   &=& - ig_V\ \epsilon\cdot V\:,
\label{3_2_1}
\end{eqnarray} 
where $k=p'-p$ is the four-momentum of the vector meson and 
$\epsilon=\epsilon(k)$ its polarisation four-vector. 

The numerical values of the $\rho$ and $\omega$ coupling constants are, 
according to \cite{Machleidt}, 
\begin{equation}
\begin{array}{rll}
\mbox{} & g_{\rho} =3.25,  \hspace{2cm} &\kappa_{\rho} = 6.1,  \\
 & g_{\omega} = 15.9,   &\kappa_{\omega} = 0.0\:.
\end{array}
\label{3_2_3}
\end{equation}

As in the Bonn potential~\cite{Bonn,Machleidt} we shall use the monopole
form factors of Eq.~(\ref{3_1_3}) with numerical values
$\Lambda_{\rho}=1.4$~GeV/c and $\Lambda_{\omega}=1.5$~GeV/c.

To take account of the $I=1$ nature of the $\rho$ meson, the vertex 
function in Eq.~(\ref{3_2_1}) must be multiplied by 
an isospin factor $\bmath{\tau}\cdot\bmath{\phi}_{\rho}$ 
in analogy to the pion-nucleon vertex.
\subsection{Pseudoscalar meson-induced $\bmath{\eta}$ production}

In the absence of polarisation data, the spin-non-flip
$\pi N\rightarrow \eta N$ amplitude is sufficient to describe the bulk
of the experimental data near threshold. In the two-component reduction
this has the structure
\begin{equation}
{\cal M}(\pi N\rightarrow \eta N)=
\bar{u}(\bmath{p}')[-ih_{\pi}]\,
\bmath{\tau}\cdot \bmath{\phi}_{\pi} \,u(\bmath{p}) \:,
\label{3_3_1}
\end{equation}
where $h_{\pi}$ may depend upon the production angle $\theta$.
For neutral pions the c.m.\ differential cross section becomes
\begin{equation}
\frac{\dd\sigma}{\dd\Omega}(\pi^0 N\rightarrow \eta N)=
\frac{p_{\eta}}{p}\ \frac{1}{16\pi^2} | h_{\pi} |^2 
\left( 1+\frac{\mu^2-m_{\pi}^2}{4M(M+\mu)} \right) \frac{1}{(1+\mu/M)^2}\:, 
\label{3_3_2}
\end{equation}
whereas for charged pions the cross section is a factor of two bigger.

Until the large Crystal Ball data set is fully analysed~\cite{Ben1}, 
the available results on the $\pi^-p\to \eta n$ differential cross section 
are generally far less systematic than those for photoproduction.
Parameterising the differential cross section as 
\begin{equation}
\label{3_3_3}
\frac{d\sigma}{d\Omega} =\frac{p_{\eta}}{p}\,[b_0\,P_0(\cos\theta)
+ b_1\,P_1(\cos\theta) + b_2\,P_2(\cos\theta)]\:, 
\end{equation}
the threshold value was determined many years ago as
$4\pi b_0=(9.1\pm 0.8)$~mb~\cite{Binnie}. This is
a little higher than the first Crystal Ball measurement at 720~MeV/c,
that gives  $4\pi b_0=(7.4\pm 0.3)$~mb~\cite{Ben2}. This group has also a
preliminary determination of the \textit{shape} of the differential cross
section at $(716\pm 16)$~MeV/c, which leads to $b_1/b_0 = -0.042$ and
$b_2/b_0 = 0.137$~\cite{Ben3}. Assuming an energy dependence of the
total cross section similar that for photoproduction shown in 
Eq.~(\ref{3_4_8}), we fix the  $\pi^0p\to \eta p$ amplitude squared to be
\begin{equation}
\label{3_3_4}
| h_{\pi}|^2 = 
[(128-303\eta^2)P_0(\cos\theta)
+23\eta\, P_1(\cos\theta) +342\eta^2\,P_2(\cos\theta)]\:\textrm{mb/sr}\:. 
\end{equation}

There are as yet no measurements of the proton analysing power near 
threshold, but the smallness of the $P_1$ coefficient in
Eq.~(\ref{3_3_4}) suggests the angular dependence arises primarily
from the interference of a small spin-non-flip $d$-wave with the dominant
$s$-wave. This could for example be due to the tail of the $D_{13}$
resonance. We therefore neglect the $P_1$ term and take
\begin{equation}
h_{\pi}= h_{\pi}^0 + \half h_{\pi}^2\left(3(\hat{\bmath{p}}_{\eta}\cdot
\hat{\bmath{k}})^2 -1\right)\:,
\label{3_3_5}
\end{equation}
where 
\begin{equation}
\label{3_3_6}
|h_{\pi}^0|^2 = (128 -303\eta^2)~\textrm{mb/sr}\,,\qquad
Re[h_{\pi}^2/h_{\pi}^0] = 1.3\eta^2\:.
\end{equation}

Multiresonance fits to projections of $\pi^-p\to \eta n$ and
$\pi^-p\to \pi^-p/\pi^0n$
data onto the $S_{11}$ channel allow one to estimate the
$\eta$-nucleon elastic scattering amplitude. Although the results,
especially on the real part, are rather sensitive to the assumptions
made~\cite{Alfred2,Tony}, a typical value of the scattering length is
$a_{\eta N} = (0.83 + 0.27i)$~fm. The elastic $\eta$-nucleon
scattering amplitude corresponding to Eq.~(\ref{3_3_1}) is
\begin{equation}
h_{\eta} = -4\pi\left(1+ \frac{\mu}{M}\right)\,a_{\eta N}\:.
\label{3_3_7}
\end{equation}

The exchanged mesons are far from the mass shell in our model.
We assume that the effects of this can be taken into account by multiplying
the pion-exchange amplitude by the monopole form factor
$F_{\pi}(q^2)$ of Eq.~(\ref{3_1_3}), and similarly for $\eta$ exchange.
\subsection{Vector-meson-induced $\bmath{\eta}$ production}
Although nothing is known directly about $\eta$ production by vector
mesons, one can use the vector-meson dominance model (VMD)  
to relate photon- and vector-meson-induced reactions~\cite{VMD}.

The isovector component of the photon couples to the $\rho$ field and
the isoscalar part to the $\omega$, though such an identification is only
valid for transversely polarised vector mesons. Neglecting higher mass
vector mesons, the photoproduction amplitude may be written in terms 
of off-shell strong-interaction amplitudes as
\begin{equation}
  {\cal M}(\gamma p\rightarrow \eta p)=
\frac{e}{f_{\rho}}\ {\cal M}_{\perp}(\rho_0 p\rightarrow \eta p) 
 + \frac{e}{f_{\omega}} {\cal M}_{\perp}(\omega p\rightarrow \eta p)\:,
 \label{3_4_1}
\end{equation}
with an analogous relation for neutron targets. 

Since universality is broken in the VMD model, we choose as effective 
$\rho$-meson coupling constant the geometric mean of the two values
derived by Benayoun \textit{et al.}~\cite{Benayoun},
\begin{equation}
\label{3_4_2a}
 \frac{f_{\rho}^2}{4\pi}= 2.4\:.
\end{equation}
This number is in reasonable agreement with those extracted from 
$\rho$-photoproduction~\cite{Bauer}.

The value of the $\omega$ coupling constant is less certain. Here
we have chosen to scale it to the $\rho$ coupling according to the 
$\Gamma(\rho\to e^+e^-)$ and $\Gamma(\omega\to e^+e^-)$ decay
widths, giving
\begin{equation} 
\frac{f_{\omega}^2}{4\pi}= 27\:.
\label{3_4_2b}
\end{equation}

Data on the photoproduction of $\eta$'s on deuterium show that near threshold
$\sigma(\gamma n \to \eta n)/\sigma(\gamma p \to \eta p)=
0.66\pm 0.03$, where some attempt has been made to include the
systematic error arising from the use a deuteron
target~\cite{Krusche}. The low cross section found for the
coherent $\gamma d\to \eta d$ reaction~\cite{Krusche} implies that
isovector photons dominate the production, so that the amplitude ratio 
is negative:
\begin{equation}
r=-\frac{{\cal M}(\gamma n\rightarrow \eta n)}
                {{\cal M}(\gamma p\rightarrow \eta p)} = 0.81\pm0.02\:.
\label{3_4_3}
\end{equation} 

The vector-meson-induced amplitudes thus become
\begin{eqnarray}
\nonumber
 {\cal M}_{\perp}(\rho_0 p\rightarrow \eta p)&=&
  \frac{f_{\rho}}{e}\left(\frac{1+r}{2}\right)\ 
      {\cal M}(\gamma p\rightarrow \eta p)\:,\\
{\cal M}_{\perp}(\omega p\rightarrow \eta p)&=& 
  \frac{f_{\omega}}{e}\left(\frac{1-r}{2}\right)\ 
      {\cal M}(\gamma p\rightarrow \eta p)\:. \label{3_4_4}
\end{eqnarray}

The threshold photoproduction amplitude is proportional to 
$\bmath{\sigma}\cdot\bmath{\epsilon}_{\gamma}$. 
The gauge-invariant photoproduction amplitude that reduces to the 
correct threshold S-wave limit is
\begin{equation}
{\cal M_{\gamma}} = R_{\gamma}\ 
 \half \bar{u}(\bmath{p}')\gamma_5
  (\not\!\epsilon \not\!k -\not\!k \not\!\epsilon)\,u(\bmath{p})
\longrightarrow 8\pi W E_{0+}\: \eta^{\,\dagger}\bmath{\sigma}\cdot
   \bmath{\epsilon}\,\zeta\:,
\label{3_4_5}
\end{equation}
where
\begin{equation}
E_{0+} =\frac{1}{8\pi W}\ \sqrt{(E+M)(E'+M)}\:(W-M)R_{\gamma}\:,
\label{3_4_6}
\end{equation}
with $k$ the photon momentum four-vector and $W$ the c.m.\ energy.

Assuming, as for the pion-induced reaction, that there is also a
contribution from the $D_{13}$ resonance as well as the $S_{11}$ at
low energies, the more general reduction is
\begin{equation}
{\cal M_{\gamma}}\longrightarrow 
8\pi W\: \eta^{\,\dagger}\left(
E_{0+}\bmath{\sigma}\cdot \bmath{\epsilon}  +\half E_{2-} 
[3\bmath{\sigma}\cdot\hat{\bmath{q}}\ \hat{\bmath{q}}\cdot\bmath{\epsilon}- 
\bmath{\sigma}\cdot \bmath{\epsilon}]
\right)\,\zeta\:,
\label{3_4_7}
\end{equation}

Taking the $d$-wave to first order, the unpolarised photoproduction
differential cross section is related to these multipoles through
\begin{equation}
\frac{\dd\sigma}{\dd\Omega}(\gamma p\rightarrow \eta p)= 
   \frac{k_{\eta}}{k} |E_{0+}|^2\,\left(1 + Re(E_{2-}/E_{0+})
\left[3(\hat{\bmath{q}}\cdot\hat{\bmath{k}})^2-1\right]\right)\:. 
\label{3_4_8}
\end{equation}

The near-threshold $\gamma p \to \eta p$ differential cross 
section~\cite{Krusche} can be parameterised~\cite{Tiator}
\begin{equation}
\label{3_4_9}
\frac{d\sigma}{d\Omega} =\frac{k_{\eta}}{k}\,[(4.59 -10.9\,\eta^2)
\,P_0(\cos\theta)-0.291\eta\,P_1(\cos\theta) 
-3.21\eta^2\,P_2(\cos\theta)]\:\mu\textrm{b/sr}\:.
\end{equation}
Neglecting the $P_1$ term, which must arise from an interference with
a small $p$-wave component, we deduce that
\begin{equation}
\label{3_4_10}
|E_{0+}|^2 = (4.59 -10.9\,\eta^2)~\mu\textrm{b/sr}\,,\qquad
Re(E_{2-}/E_{0+})=-0.70\eta^2\:.
\end{equation}
At threshold 
$|E_{0+}|= 0.0214$~fm and hence $|R_{\gamma}|= 0.146$~fm$^2$.
The energy variation of $E_{0+}$ is mainly a reflection 
of the dominance of the $N^*(1535)$ resonance near the $\eta$
threshold.

In the same approximation, the proton analysing power is~\cite{Tiator}
\begin{equation}
\label{3_4_11}
A_y=-3\, Im(E_{2-}/E_{0+})\sin\theta\cos\theta\:.
\end{equation} 
$A_y$ remains small for photon
energies below 850~MeV~\cite{Bock}, with evidence for both $sp$ and
$sd$ interference. This gives
\begin{equation}
\label{3_4_12}
Im(E_{2-}/E_{0+}) = (-3\pm0.5)\eta^2 + (16\pm 4)\eta^4\:.
\end{equation}

We now make the \textit{ad hoc} assumption that all three polarisation
states have the same strength in the vector-meson-induced reaction.
The amplitude is then proportional to
$\bmath{\sigma}\cdot\bmath{\epsilon}_V$, with the proportionality 
constant being determined by the vector-meson-dominance model.
The relativistic form which has this limit is
\begin{eqnarray}
{\cal M}(Vp\rightarrow\eta p)& = & 
   R_V  \bar{u} (\bmath{p}')\left[ \half \gamma_5
  (\not\!\epsilon \not\!k -\not\!k \not\!\epsilon) + 
   m_V  \gamma_5 \not\!\epsilon\,\right] u(\bmath{p})\nonumber \\
  &=&\sqrt{(E+M)(E'+M)}\:(W-M+m_V) R_V\:\eta^{\,\dagger}\bmath{\sigma}\cdot
   \bmath{\epsilon}\, \zeta\:.
\label{3_4_13}
\end{eqnarray}
The scale factors $R_V$ are determined from the VMD model, 
Eqs.~(\ref{3_4_4}), evaluated at the $\eta$ threshold 
$W=M+\mu$:
\begin{equation}
R_{\rho}= R_{\gamma}\frac{f_{\rho}}{e}\left(\frac{1+r}{2}\right)\,
     \left[ \frac{\mu}{\mu+m_{\rho}} \right]\:,\hspace{1cm}
R_{\omega}= R_{\gamma} \frac{f_{\omega}}{e}\left(\frac{1-r}{2}\right)\,
  \left[ \frac{\mu}{\mu+m_{\omega}} \right]\,\cdot
\label{3_4_14}
\end{equation}
From the value of $R_{\gamma}$, we deduce that 
$|R_{\rho}|= 0.916$~fm$^2$ and $|R_{\omega}|= 0.310$~fm$^2$
at threshold.

As before, we attempt to describe the off-shell extrapolation through a form 
factor. Since we are starting from data with zero-mass photons, the correct
form factor at the $\rho$-induced photoproduction vertex is
\begin{equation}
\tilde{F}_{\rho}(q^2)= F_{\rho}(q^2)/F_{\rho}(0)\:.
\label{3_4_17}
\end{equation}
We take for $F_{\rho}(q^2)$ the monopole form factor of Eq.~(\ref{3_1_3}).
This is not in contradiction with the off-shell behaviour of the 
$\eta$-photoproduction amplitude measured \textit{via} electron
scattering at modest $q^2$~\cite{Thompson}.
\subsection{Deuteron vertex}
Both nucleons are close to their mass shells at the deuteron vertex, so that 
each pole gives a contribution after integration over 
the internal energy variable. The deuteron vertex matrix element 
corresponding to the deuteron S- and D-state wave functions $\phi_S(r)$ 
and $\phi_D(r)$ is
\newpage
\[
iS_F^c(k)\,(-i\Gamma\cdot\epsilon_d^{\star})\,iS_F(p_d-k)=\]
\[u_c(\bmath{k})\, 
(2\pi)^{3/2}\:\eta_c^{\dagger}\, \frac{-1}{\sqrt{2}}\left[
  \bmath{\sigma}\cdot\bmath{\epsilon}_d^{\dagger}\ \phi_S (\bmath{Q}_R)
  +\left\{ \bmath{\sigma}\cdot\bmath{\epsilon}_d^{\dagger} - 
    3\bmath{\sigma}\cdot\hat{\bmath Q}_R\ 
\bmath{\epsilon}_d^{\dagger}\cdot\hat{\bmath Q}_R
  \right\}\phi_D (\bmath{Q}_R) \right]\zeta\:
\bar{u}(\bmath{p}_d-\bmath{k})
\]
\begin{equation}
 \times \frac{\sqrt{2M_d}}{2M} \:(2\pi)
         \{\delta(k^0-E_n)+\delta(p_d^0-k^0-E_p)\}\:.
\label{3_6_1} 
\end{equation}
$E_n$ and $E_p$ are the nucleon on-shell energies so that, \textit{e.g.}\ 
$E_n=\sqrt{\bmath{k}^2+M^2}$, and $\bmath{Q}_R$ is the  
proton-neutron relative momentum.\\
\newpage
\section{Amplitude for $\bmath{pp\rightarrow pp\eta}$}

We describe the reaction $pp\rightarrow pp\eta$ in terms of Feynman
diagrams, the prototype of which is displayed in Fig.~1, considering
the four exchanges, $\rho$, $\omega$, $\pi$, and $\eta$. The 
matrix elements are evaluated at threshold, where the momenta in the 
final state  vanish and the values of the corresponding incident proton 
energies and momenta are given in Eq.~(\ref{2_1}). We first study 
neutral-particle exchanges; the necessary isospin factors will be
derived at the end of the section.

\noindent
\input epsf
\begin{figure}[ht]
\begin{center}
\epsffile[80 615 480 735]{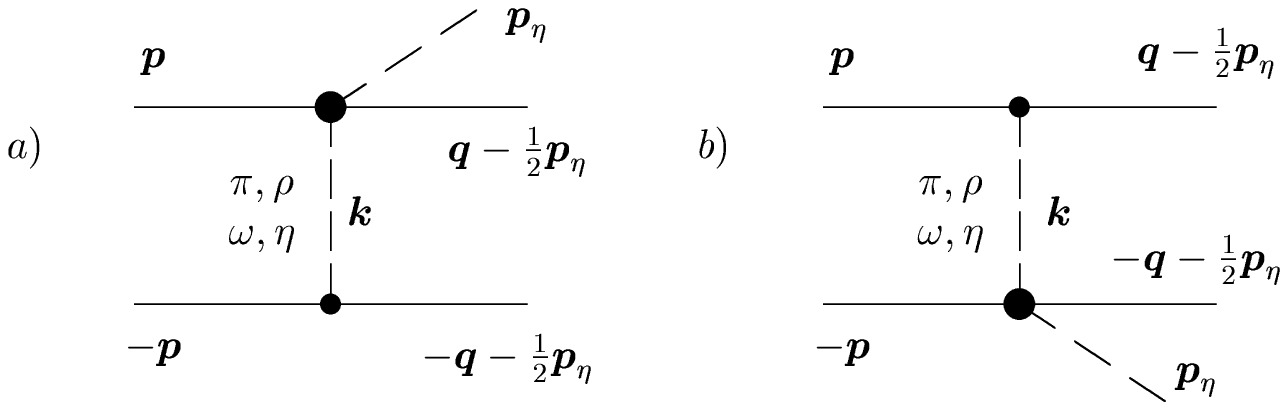}
\caption{Meson exchange diagrams for $pp\rightarrow pp\eta$. Diagrams 
$c)$ and $d)$ are obtained from $a)$ and $b)$ by interchanging the 
final state nucleons.}
\end{center}
\label{fig1}
\end{figure}
\subsection{Vector-exchange diagrams}

The propagator for vector meson exchange is
\begin{equation}
iD_{\mu\nu}(k,m)=\frac{i(-g_{\mu\nu}+k_{\mu}k_{\nu})}{k^2-m_V^2}\:\cdot
\label{4_1_1}
\end{equation}
When the vector meson four-momentum  $k_{\mu}$ is contracted into 
the $VNN$ vertex, it vanishes identically. We thus obtain for 
$\rho^0$ exchange in diagram $a$
\begin{equation}
{\cal M}_a(pp\rightarrow pp\eta) =
{\cal M}_{\mu}(p\rightarrow p\rho^0) \frac{-i}{k^2-m_{\rho}^2}
{\cal M}^{\mu}(\rho^0p\rightarrow \eta p)\:.
\label{4_1_2}
\end{equation}
Using the expressions for the matrix elements of the vertex $\rho pp$ 
and amplitude $\rho p\rightarrow \eta p$,
given in Eqs.~(\ref{3_2_1}) and (\ref{3_4_12}), leads to the threshold
expression 
\begin{equation}
{\cal M}_a(pp\rightarrow pp\eta) = 4M
\frac{g_{\rho}R_{\rho}}{M\mu+m_{\rho}^2}
\left[F_{\rho}^2(-M\mu)/F_{\rho}(0)\right] {\cal K}_a\:,
\label{4_1_3}
\end{equation}
with 
\begin{equation}
{\cal K}_a =
 a\: [\eta_3^{\,\dagger}\,\bmath{\sigma}\cdot\bmath{p}\,\zeta_1]\,[
    \eta^{\,\dagger}_4\,\zeta_2]
-ib\: [\eta_3^{\,\dagger}\,\bmath{\sigma}\,\zeta_1]\cdot
    [\eta_4^{\,\dagger}\,\bmath{p}\times\bmath{\sigma}\,\zeta_2]\:.
\label{4_1_4}
\end{equation}
The parameters $a$ and $b$ are given by
\begin{eqnarray}
\nonumber
 a &=& (M-m_{\rho})(1-\frac{\mu\kappa_{\rho}}{4M})\:, 
\\
 b &=& \half (m_{\rho}+\mu)(1+\kappa_{\rho})\:.
   \label{4_1_5}
\end{eqnarray}
The momentum of the intermediate vector meson is $\bmath{k}=-\bmath{p}$ 
and the form factor is evaluated at mass squared $k^2=-M\mu$.

We can similarly write down the expression for diagram $b$, where 
the two vertices are interchanged:
\begin{equation}
{\cal K}_b=
 - a\:[\eta_3^{\,\dagger}\,\zeta_1]\,[
    \eta^{\,\dagger}_4\, \bmath{\sigma}\cdot\bmath{p}\,\zeta_2]
 +ib\:[\eta_3^{\,\dagger}\,\bmath{p}\times\bmath{\sigma}\,\zeta_1]\cdot
    [\eta_4^{\,\dagger}\,\bmath{\sigma}\,\zeta_2]\:,
\label{4_1_6}
\end{equation}
with $\bmath{k}=+\bmath{p}$.

It is easy to interpret the $b$-terms if we explicitly write the sum over
the polarisation states of the $\rho$ meson. Thus, in Eq.~(\ref{4_1_4}),
this gives
\begin{equation}
 [\eta_3^{\,\dagger}\,\bmath{\sigma}\,\zeta_1]\,
    [\eta_4^{\,\dagger}\,\bmath{p}\times\bmath{\sigma}\,\zeta_2]=
 \sum\:[\eta_3^{\,\dagger}\,\bmath{\sigma}\cdot 
    \bmath{\epsilon}_{\rho}\,\zeta_1]\,
 [\eta_4^{\,\dagger}\, \bmath{\epsilon}_{\rho}\cdot 
(\bmath{p}\times\bmath{\sigma})\,\zeta_2]
\label{4_1_7}
\end{equation}
Here we recognise the vertex for the $\rho NN$ coupling as 
$\bmath{\epsilon}_{\rho}\cdot(\bmath{k}\times\bmath{\sigma})$
and that for $\rho$-induced $\eta$ production as 
$\bmath{\sigma}\cdot\bmath{\epsilon}_{\rho}\,$. The small $a$-term
arises from the part of the $\rho NN$ coupling that contains the
vertex factor $\bmath{\epsilon}_{\rho}\cdot\bmath{k}\,$.

At threshold the initial total spin of the $pp$ system is $S=1$
with $S=0$ in the final state and we shall let $\bmath{\epsilon}$
denote the spin vector of the spin-one $pp$ state and $\eta$ the
spin-zero state vector. The contribution of diagrams $a$ and $b$
is the same as that for the crossed diagrams $c$ and $d$:
\begin{equation}
{\cal K}_a + {\cal K}_b = {\cal K}_c + {\cal K}_d
=- 2(2b-a)\ \bmath{p}\cdot\bmath{\epsilon}_i\ \eta_f\:.
\label{4_1_9}
\end{equation}

From Eq.~(\ref{4_1_3}) and (\ref{4_1_9}), 
the total $\rho^0$ contribution becomes
\begin{equation}
{\cal M}_{\rho}(pp\rightarrow pp\eta) =
({\cal A}_{\rho} -2{\cal B}_{\rho})\
\bmath{p}\cdot\bmath{\epsilon}_i \ \eta_f\:,
\label{4_1_10}
\end{equation}
where 
\begin{eqnarray}
\nonumber
{\cal A}_{\rho} &=& 16M
\frac{g_{\rho}R_{\rho}}{M\mu+m_{\rho}^2}
\left[F_{\rho}^2(-M\mu)/F_{\rho}(0)\right] a\:, \\
{\cal B}_{\rho} &=& 16M
\frac{g_{\rho}R_{\rho}}{M\mu+m_{\rho}^2}
\left[F_{\rho}^2(-M\mu)/F_{\rho}(0)\right]b\:.
\label{4_1_11}
\end{eqnarray}
With the values of the  parameters given in section~3, it can be seen that
the ${\cal A}$-term is of no numerical importance for $\rho$ exchange.

When the $d$-waves are included in the dominant $b$-term, the amplitude of
Eq.~(\ref{4_1_10}) is modified slightly to read
\begin{equation}
{\cal M}_{\rho}(pp\rightarrow pp\eta) =
({\cal A}_{\rho} -2{\cal B}_{\rho})\:
\left[\bmath{p}\cdot\bmath{\epsilon}_i
+\half (E_{2-}/E_{0+})(3\bmath{p}\cdot\hat{\bmath{q}}\:
\hat{\bmath{q}}\cdot\bmath{\epsilon}_i -
\bmath{p}\cdot\bmath{\epsilon}_i ) \right]\: \eta_f\:.
\label{4_1_12}
\end{equation}
To first order in $E_{2-}/E_{0+}$, this gives exactly the same form
as in the $\gamma p \to \eta\,p$ cross section of Eq.~(\ref{3_4_8}).

The amplitudes for $\omega$ are completely analogous to those for
$\rho$ exchange, so that at threshold
\begin{equation}
{\cal M}_{\omega}(pp\rightarrow pp\eta) =
({\cal A}_{\omega} -2{\cal B}_{\omega})\
\bmath{p}\cdot\bmath{\epsilon}_i \ \eta_f\:,
\label{4_1_13}
\end{equation}
where the amplitudes are as defined in Eq.~(\ref{4_1_11}) and the parameters
in Eq.~(\ref{4_1_5}). 
\subsection{Pseudoscalar-exchange diagrams}
The four diagrams with $\pi^0$ exchange are labelled as in Fig.~1. 
The contribution from diagram $a$ is
\begin{equation}
{\cal M}_a(pp\rightarrow pp\eta) =
{\cal M}(p\rightarrow p\pi^0)\, \frac{i}{k^2-m_{\pi}^2}\,
{\cal M}(\pi^0p\rightarrow \eta p)\:.
\label{4_2_1}
\end{equation}
Inserting the forms for the $\pi pp$ vertex  and $\eta$-production amplitude,  
given by Eq.~(\ref{3_1_1}) and (\ref{3_3_1}), into Eq.~(\ref{4_2_1}) yields
\begin{equation}
{\cal M}_a(pp\rightarrow pp\eta) =
2MG_{\pi}\,h_{\pi}\ \frac{1}{M\mu+m_{\pi}^2}
\left[F_{\pi}^{\,2}(-M\mu)\right]\  {\cal K}_a\:,
\label{4_2_2}
\end{equation}
where
\begin{equation}
{\cal K}_a= [\eta_3^{\,\dagger}\,\zeta_1]\,
  [\eta^{\,\dagger}_4\,\bmath{\sigma}\cdot\bmath{p}\,\zeta_2]\:.
\label{4_2_3}
\end{equation}
Using the same techniques as for $\rho$ exchange, the sum of the 
amplitudes for diagrams $c$ and $d$ is the same as that for $a$ and $b$:
\begin{equation}
{\cal K}_a+{\cal K}_b= {\cal K}_c+{\cal K}_d= 
2\ \bmath{p}\cdot\bmath{\epsilon}_i \ \eta_f\:.
\label{4_2_4}
\end{equation}
The final result is then
\begin{equation}
{\cal M}_{\pi}(pp\rightarrow pp\eta) =  {\cal D}_{\pi}
 \ \bmath{p}\cdot\bmath{\epsilon}_i \ \eta_f\:,
\label{4_2_5}
\end{equation}
where
\begin{equation}
{\cal D}_{\pi} = 8MG_{\pi}\,h_{\pi}\ \frac{1}{M\mu+m_{\pi}^2}
\left[F_{\pi}^{\,2}(-M\mu)\right]\:.
\label{4_2_6}
\end{equation}
With the inclusion of $d$-waves, the input amplitude $h_{\pi}$ has an
angular dependence given by Eq.~(\ref{3_3_5}).

The expressions for the $\eta$-exchange amplitudes are algebraically 
identical to those of pion exchange, with the $\pi$ index being replaced by
an $\eta$. Of course, as discussed in Section~2, poor knowledge of the value of
the $\eta NN$ coupling constant makes the strength of any $\eta$
contribution extremely uncertain.
\subsection{Isospin considerations}
The expressions derived for the various amplitudes correspond to
neutral-meson exchange for the $pp$ case. The $NN \rightarrow NN\eta$ 
isospin-one production amplitude is 
\begin{equation}
{\cal M}_1= {\eta}_f^{\,\dagger}\  \bmath{p} \cdot \bmath{\epsilon}_i \ 
        \left[({\cal A}_{\rho}-2{\cal B}_{\rho})+
({\cal A}_{\omega}-2{\cal B}_{\omega})+ {\cal D}_{\pi}+{\cal D}_{\eta}\right]\ 
      \bmath{\chi}_f^{\,\,\dagger}\cdot\bmath{\chi}_i\:.  
\label{4_3_1}
\end{equation}  
It is straightforward to generalise this to the case of 
isospin zero~\cite{GW}:
\begin{equation}
{\cal M}_0= \bmath{p} \cdot \bmath{\epsilon}_f^{\,\,\dagger}\  {\eta}_i\ 
 \left[-3({\cal A}_{\rho}+2{\cal B}_{\rho})+
({\cal A}_{\omega}+2{\cal B}_{\omega}) 
-3{\cal D}_{\pi}+{\cal D}_{\eta}\right]\ \phi_f^{\,\dagger}\, \phi_i \ .
\label{4_3_2}
\end{equation}
Here $\bmath{\chi}$ and $\phi$ are isospin-1 and -0 operators and
the subscripts refer to the particles exchanged.

The spin-average of the square of the matrix element becomes
\begin{equation}
\fourth \sum \left|{\cal{M}}(pp\rightarrow pp\eta)\right|^{2}=
 \frac{p^2}{4} 
\left|({\cal A}_{\rho}-2{\cal B}_{\rho})+
({\cal A}_{\omega}-2{\cal B}_{\omega})+ 
{\cal D}_{\pi}+{\cal D}_{\eta}\right|^2
\label{4_3_3}
\end{equation}
for the $pp$ case and
\begin{eqnarray}
\fourth \sum \left| {\cal{M}}(pn\rightarrow np\eta) \right| ^{2} &=&
 \frac{p^2}{16}  \left[
  \left|-3({\cal A}_{\rho}+2{\cal B}_{\rho})+
  ({\cal A}_{\omega}+2{\cal B}_{\omega}) 
-3{\cal D}_{\pi}+{\cal D}_{\eta}\right|^2\right.
 \nonumber \\  \nonumber \\
&+& \left.\left|({\cal A}_{\rho}-2{\cal B}_{\rho})+
   ({\cal A}_{\omega}-2{\cal B}_{\omega})+ 
{\cal D}_{\pi}+{\cal D}_{\eta}\right|^2\right]
\label{4_3_4}
\end{eqnarray}
for $pn$. Application of the formulae in Section~2 then give the 
cross sections.
\subsection{Kinematics}

Since we are only considering cases where the excess energy is low,
the transformation between the final $\eta\,N$ and $\eta\,NN$ systems
is non-relativistic. Taking the laboratory momentum to be the same
in the $\eta\,N$ and $\eta\,NN$ cases, one sees that the c.m.\
momentum is lower in the single-nucleon than the two-nucleon case:
\begin{equation}
\label{4_4_1}
\bmath{p}_{\eta N}= \left(\frac{1+\mu/2M}{1+\mu/M}\right)\,
\bmath{p}_{\eta NN}\:.
\end{equation}
\newpage
\section{Initial and Final Nucleon-Nucleon \mbox{Distortion}}

Most of the energy dependence observed in the cross sections for the
$NN\to NN\eta$ reactions near threshold can be ascribed to the behaviour
of the three-body phase space that has been modified by the very strong
nucleon-nucleon final state interaction (fsi). It has been traditional to
incorporate the effects of such an fsi by multiplying the predicted cross
section by a Watson enhancement factor. This prescription gives a rapid energy
variation parameterised in terms of the $NN$ scattering length and effective
range~\cite{Gold}. It does, however, give too steep a fall at higher
$NN$ relative momenta~\cite{FW2} and does not attempt to provide an overall
normalisation factor.

In their analysis of the experimental data, the authors of \cite{Calen4}
took the enhancement factor as the ratio of the
squares of the interacting $NN$ $S$-state wave function to the
equivalent plane wave. In the absence of a model for the
$\eta$-production operator, they evaluated the ratio at a fixed
radius of $r=1$~fm, which is close to the maximum of the $NN$
density~\cite{Paris}.  

To obtain more realistic enhancement factors, we must study further
the meson-production operator.
In the plane-wave approximation, the dominant $\rho$-exchange 
contribution from the diagram of Fig.~1a can be described by a potential 
\begin{equation}
 V_{\rho}(\bmath{r}) \propto (\bmath{\sigma}_1\times \bmath{\sigma}_2)
\cdot\hat{\bmath{r}}\: \left( 1 + \frac{1}{m_{\rho}^{*}r}\right) 
\frac{e^{-m_{\rho}^{*}r}}{m_{\rho}^{*}r}\:\cdot
\label{5_1}
\end{equation}

In the vicinity of the $\eta$ threshold, the energy transfer is shared
equally by the two nucleons, so that the range of this propagation is
determined by the reduced $\rho$-mass, given by
\begin{equation}                                                     
\label{5_2}
m_{\rho}^{*\,2} = m_{\rho}^{2}-\omega^{2}/4\:,
\end{equation}                                                     
where $\omega$ is the total energy of the $\eta$ meson.

Taking matrix elements of the potential between interacting nucleon
but $\eta$ plane waves and projecting out the threshold angular momentum 
transition, we are left with the amplitude
\begin{equation}
\label{5_3}
{\cal M}_{\rho}^{int} ={\cal M}_{\rho} C_{\rho}
   \:\int_0^{\infty} \psi_{k}^{(-)\,*}(r)\,
Y_1(m_{\rho}^{*},r)\,j_0(\half p_{\eta}r)
\,\psi_{p}^{(+)}(r)\, r^2\,dr\:,
\end{equation}
with ${\cal M}_{\rho}$  the appropriate plane wave $\rho$-exchange 
amplitude of Eq.\ (\ref{4_3_1}) or (\ref{4_3_2}), and where
\begin{equation}
\label{5_4}
C_{\rho} =\frac{1}{p}m_{\rho}^{*\,2}(m_{\rho}^{*\,2}+p^2)
\end{equation}
and
\begin{equation}
\label{5_5}
Y_1(m_{\rho}^{*},r)=\left( 1 + \frac{1}{m_{\rho}^{*}r}\right)\,
\frac{e^{-m_{\rho}^{*}r}}{m_{\rho}^{*}r}\:.
\end{equation}
The wave function $\psi_{k}(r)$ describes the $S$-wave of the final
$I=1$, $J=0$ or $I=0$, $J=1$ $NN$ system and $\psi_{p}(r)$  
the incident high energy $NN$ $P$-waves. 

The exchange of other negative parity mesons, $(\omega, \pi, \eta)$,
leads to a very similar structure, though with different effective
masses given by Eq.~(\ref{5_2}). 

When monopole form factors of the type given in Eq.~(\ref{3_1_3}) are
introduced at the $\rho NN$ and production vertices, the only modification
to the formula is to replace the $Y_1(m_{\rho}^{*},r)$ propagator in 
Eq.~(\ref{5_5}) by
\begin{equation}
\label{5_6}
\bar{Y}_1(m_{\rho}^{*},r)= 
\frac{(\Lambda^{*\,2}+p^2)^2}{(\Lambda^{*\,2}-m_{\rho}^{*\,2})^2}\,
\left[Y_1(m_{\rho}^{*},r)-\frac{1}{2m_{\rho}^{*\,2}}
\left(\frac{2}{r^2}+\frac{2\Lambda^*}{r}
+(\Lambda^{*\,2}-m_{\rho}^{*\,2})\right)\,e^{-\Lambda^* r}\right]\:,
\end{equation}
with
\begin{equation}
\label{5_6a}
\Lambda^{*\,2}=\Lambda^2-\omega^2/4
\end{equation}

For $r \ll 1/\Lambda^*$, $\bar{Y}_1(m_{\rho}^{*},r) = O(r)$ and so the 
form factor reduces a little the sensitivity to the small-$r$
components of the wave function.

In order to investigate purely the effect of the $NN$ fsi near threshold,
$\psi_{k}(r)$ is replaced by the plane wave spherical Bessel function
$j_1(pr)$. Near threshold the $\eta$ plane wave factor
$j_0(\half p_{\eta}r)$ can be taken to be unity. We then quantify the final
state enhancement through the ratio of the amplitudes
\begin{equation}
\label{5_7}
E_x(k) = {\cal M}_x^{\textrm{\scriptsize int}}/
{\cal M}_x^{\textrm{\scriptsize pw}}
\end{equation}
calculated from Eq.~(\ref{5_3}) with interacting and plane $NN$
final-state waves.

The enhancement factors $|E_x|^2$ are shown in Fig.~2 for both 
$I=0$ and $I=1$ $NN$ states in the case of $\rho$-exchange. It can be seen
that Coulomb effects are significant for $pp$ final states with
$k\leq 0.2$~fm$^{-1}$. For $\pi$-exchange the energy dependence is almost
identical, though the overall magnitude is somewhat less. To illustrate this,
the $I=0$ enhancement factor for $\pi$-exchange is shown multiplied by
a factor of $1.85$.

Because the pole of the antibound state in the $pp$ system is closer to
threshold than that of the deuteron in $np$, there is significant
energy variation in the ratio of the ($I$=1)/($I$=0) enhancement factors.
Nevertheless, as can be seen in Fig.~3, for $k^2 > 0.2$~fm$^{-2}$ the
ratio is roughly constant at 1.85. This deviation from unity is
important in the understanding of the $pp\to pp\eta/pn\to np\eta$
ratio. The empirical approach used in \cite{Calen4}, where the wave
function squared is evaluated at a fixed radius of 1~fm, would lead to
a ratio with a similar energy dependence to that shown in 
Fig.~3 but lower in magnitude by an overall factor of $0.8$

\noindent
\input epsf
\begin{figure}[ht]
\begin{center}
\mbox{\epsfxsize=3.5in \epsfbox{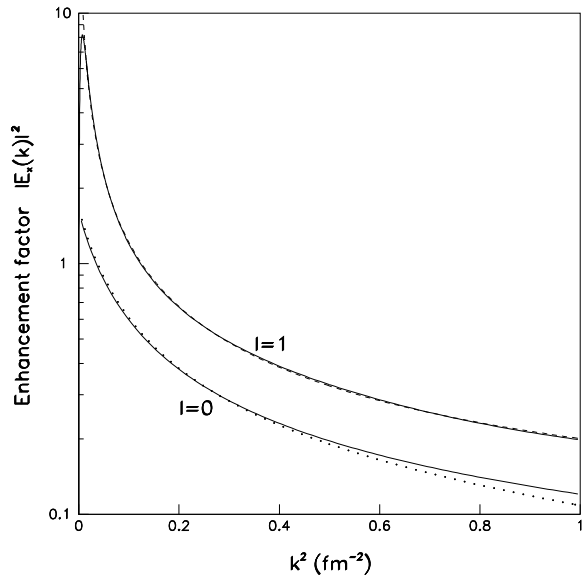}}
\caption{Enhancement factors, defined by Eq.~(\protect\ref{5_7}), evaluated
for isospin-one and -zero $NN$ $S$-waves using Paris wave 
functions~\protect\cite{Paris}.
The solid lines are without the Coulomb repulsion, whereas the broken
line includes this effect in the $I=1$ case. The calculations were done
for $\rho$-exchange but $\pi$-exchange leads to a very similar energy
dependence though reduced in magnitude by a factor of about $1.85$ in the
$I=0$ case and $1.65$ for $I=1$. This is illustrated for $I=0$ by the
dotted curve that has been scaled up by $1.85$.}
\end{center}
\label{fig2}
\end{figure}

Though the initial-state interaction is expected to vary little with
energy, the evaluation of an $NN$ wave function from a potential at
energies as high as 1300-1400~MeV is extremely dubious. In 
order to take the initial flux damping into account, the amplitudes 
are multiplied by the factor $\eta_L=e^{-Im\{\delta_L\}}$. This is 
typically $0.77$ for the $^3$P$_0$ state and $0.73$ for $^1$P$_1$.

\noindent
\input epsf
\begin{figure}[ht]
\begin{center}
\mbox{\epsfxsize=3.5in \epsfbox{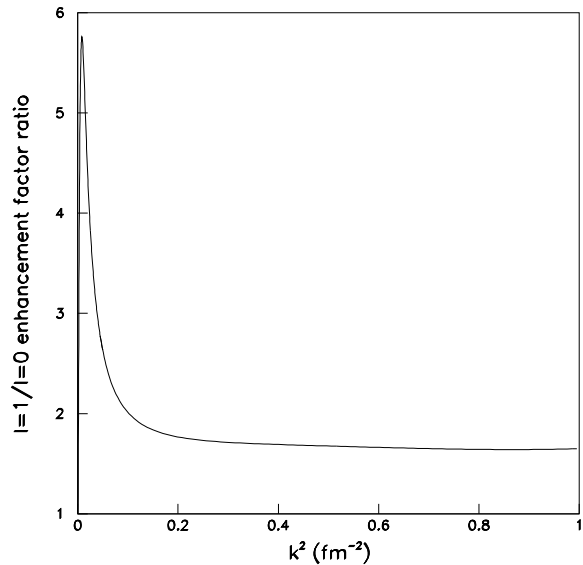}}
\caption{Ratio of enhancement factors for isospin-one and -zero $NN$
final states evaluated for $\rho$-exchange using Paris wave
functions with Coulomb effects~\protect\cite{Paris}. The ratio of the 
squares of
the wave functions at $r\approx 1$~fm is similar in shape but reduced in
magnitude by about 20\%.}
\end{center}
\label{fig3}
\end{figure}

We have neglected the distortion of the final $\eta$ wave. This is, in
part, due to the difficulty of including consistently final state
interactions simultaneously in the $pp$ and $\eta p$ channels. There is,
furthermore, a lack of a realistic $\eta$-nucleon potential in the
literature. On the basis of potential-model fits to $\eta$-nucleon
scattering parameters, Garcilazo and Pe\~{n}a predict a 
strong threshold enhancement in $\eta$-deuteron elastic
scattering~\cite{Teresa}. 
However, probably as an artefact of their one-term separable assumption,
their potentials are extremely attractive at short distances
($r < 0.2$~fm). Since $\eta$-production is sensitive to wave functions
at small $r$, using these potentials as a basis for estimating final 
state interaction effects can lead to very unrealistic values. We
have therefore represented the $\eta$ by a plane wave though, as we shall 
see later, there are indications from the production data that this 
is inadequate.
\newpage
\section{Amplitude for $\bmath{pn\to d\eta}$}

The triangle diagrams describing the $pn\rightarrow d\eta$ reaction
are shown in Fig.~4. They are very similar to those for $pn\to np\eta$ 
with the addition of a neutron-proton final-state interaction to
produce the deuteron.
\noindent
\input epsf
\begin{figure}[ht]
\begin{center}
\epsffile[96 602 518 740]{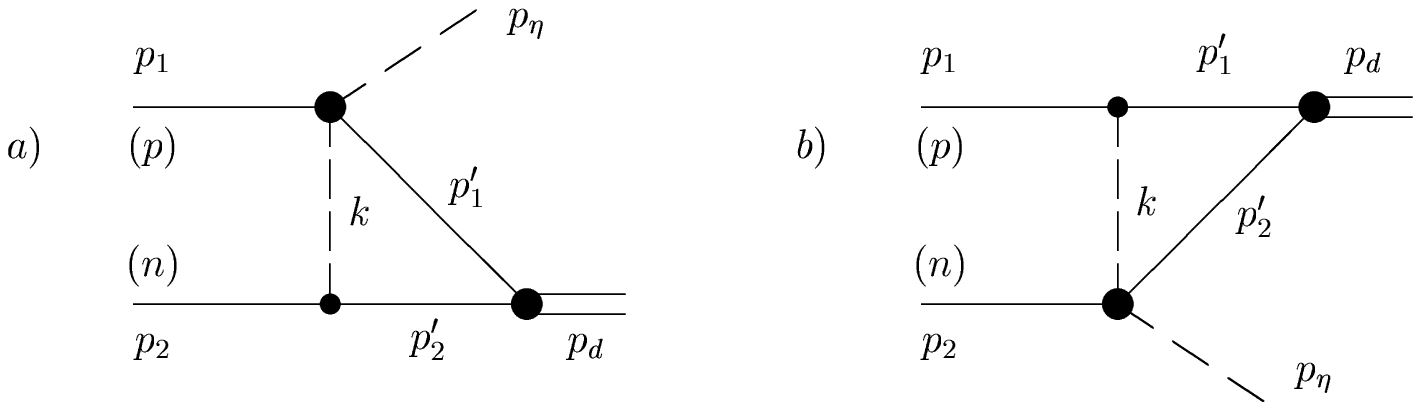}
\caption{Meson exchange diagrams for $pn\rightarrow d\eta$.}
\end{center}
\label{fig4}
\end{figure}

The Feynman amplitude corresponding to Fig.~4a for $\rho$ exchange is
\[
{\cal M}_{\rho}=
 \frac{3}{\sqrt{2}}\int\frac{d^4k}{(2\pi)^4}
  \frac{-i}{k^2-m_{\rho}^2}\:
{\bar{u}}_c(p_2)\:{\cal M}_{\mu}(n_c\rightarrow \rho^0n_c)\]
\begin{equation}
\times\: iS_F^c(p_2')\,(-i\Gamma\cdot\epsilon_d^{\star})\,iS_F(p_1')
\:{\cal M}^{\mu}(\rho^0p\rightarrow \eta p)\,u(p_1)\:,
\label{6_1}
\end{equation}
where we have treated the incident neutron in the charge conjugate 
representation. Here $k$ is the four-momentum of the $\rho$-meson
and we can neglect the term proportional to $k_{\mu}k_{\nu}$
in the $\rho$ propagator since this gives only binding energy
contributions. The isospin factor of $3/\sqrt{2}$ takes into
account the contribution of charged intermediate mesons. There is,
in addition, the diagram of Fig.~4b where the $\eta$ production takes place at
the neutron vertex.

We calculate the triangle diagram in the spectator approximation 
by performing the integration over $q^0$, after suitably closing the 
contour of integration. This means exploiting formula of
Eq.~(\ref{3_6_1}). After some straightforward manipulations we get in
configuration space, as the sum of both $\rho$-exchange diagrams,
\[
{\cal M}_{\rho}=-\frac{3}{\sqrt{2}} ({\cal A}_{\rho}+2{\cal B}_{\rho})
\bmath{p}\cdot\bmath{\epsilon}_d^{\ \dagger}\ {\eta}_i\, 
\frac{\sqrt{2M_d}}{2M}\, \]
\begin{equation}
\label{6_2}
 \times\:C_{\rho}\:\int_0^{\infty}
\left[\psi_{S}(r)+2\psi_{D}(r)\right]\,
\bar{Y}_1(m_{\rho}^{*},r)\,j_0(\half p_{\eta}r)\,
\psi_{p}^{(+)}(r)\, r^2\,dr\:.
\end{equation}
where $\bar{Y}_1(m_{\rho}^{*},r)$ and $C_{\rho}$  are defined 
in Eqs.~(\ref{5_6}) and (\ref{5_4}).

In a similar fashion, we find the pion-exchange amplitude to be
\[
{\cal M}_{\pi}=-\frac{3}{\sqrt{2}} {\cal D}_{\pi}
\bmath{p}\cdot\bmath{\epsilon}_d^{\ \dagger}\ {\eta}_i\,
\frac{\sqrt{2M_d}}{2M}\, \]
\begin{equation}
\label{6_3}
 \times\:C_{\pi}\:\int_0^{\infty}
\left[\psi_{S}(r)+2\psi_{D}(r)\right]\,
\bar{Y}_1(m_{\pi}^{*},r)\,j_0(\half p_{\eta}r)\,\psi_{p}^{(+)}(r)
\end{equation}
with $\bar{Y}_1(m_{\pi}^{*},r)$ and $C_{\pi}$ being defined as in 
Eqs.~(\ref{5_6}) and (\ref{5_4}), but with $m_{\rho}$ replaced 
by $m_{\pi}$.  

The contributions from $\omega$ and $\eta$ exchange can be derived
immediately from Eqs.~(\ref{6_2}) and (\ref{6_3}) in an
analogous manner to that for $pp\to pp\eta$.

The $\eta$-production operator, especially that part corresponding to
$\rho$ exchange, is of short range. At short distances and low
energies the deuteron $S$-state wave function and the $I=0$ scattering
wave functions at low energies are related by the extrapolation 
theorem~\cite{FW3}. 
\begin{equation}
\label{6_4}
\psi_{k}^{(+)}(r)\approx
-\frac{1}{\sqrt{2\alpha(k^2+\alpha^2)}}\,\psi_S(r)\,e^{i\delta_s}\:,
\end{equation}
where $\delta_s$ is the $S$-wave triplet phase shift. If we neglect the
$S$-$D$ coupling, the relation becomes exact as
$k^2\to -\alpha^2=-M\,\varepsilon_d$, where $\varepsilon_d$
is the deuteron binding energy. 

By comparing Eq.~(\ref{6_3}) with Eq.~(\ref{5_4}), corrected for
isospin and form factor effects, the extrapolation theorem allows us 
to estimate the triplet contribution to the $pn\to pn\eta$ cross 
section in terms of that for $pn\to d\eta$ independent of the
details of the meson exchanges responsible. Generalising the result of
Ref.~\cite{FW2}, if we parameterise the low energy
deuteron cross section estimates as
\begin{equation} 
\label{6_5}
\sigma_{pn\to d\eta}(Q)\approx a\sqrt{Q}\,(1+b\,Q)\:,
\end{equation}
then the integration over phase space leads to the prediction that
\begin{equation}
\label{6_6}
\sigma_{pn\to pn\eta}^{I=0}(Q)\approx 
\frac{1}{4}x^{3/2}\,\sqrt{Q}\,
\left(1+\sqrt{1+x}\,\right)^{\!-2}\,a\,
\left[1+\frac{b}{2}\,Q\left(1+\frac{1}{2}\,\frac{x}{(1+\sqrt{1+x}\,)^2}
\right)\right]\:,
\end{equation}
where $x=Q/\varepsilon_d$.

Deviations from the predictions of Eq.~(\ref{6_5}) in our results will
mainly arise from corrections to the approximation of Eq.~(\ref{6_4})
in the scattering domain and, in particular, from $D$-state effects.
\newpage
\section{Results}

Since the value of the $\eta$-nucleon coupling constant is very poorly known,
we shall start by neglecting $\eta$ exchange, to discuss it later.
One would expect, on the basis of Eqs.~(\ref{4_3_3}) and (\ref{4_3_4}), that
with pure $\rho$ or $\pi$ exchange the $pn\!:\!pp$ cross section ratio would
be a factor of 5, which is not far from the observed value of
$6.5\pm 1.0$~\cite{Calen5}. This agreement is, however, destroyed by
the final-state interaction, which gives a bigger enhancement for the 
$I=1$ state than for the $I=0$ (see Fig.~2). Pure $\rho$ or $\pi$ 
exchange would then predict a ratio of 3 or even less. 

\vspace{-3mm}
\noindent
\input epsf
\begin{figure}[hbt]
\begin{center}
\mbox{\epsfxsize=3.5in \epsfbox{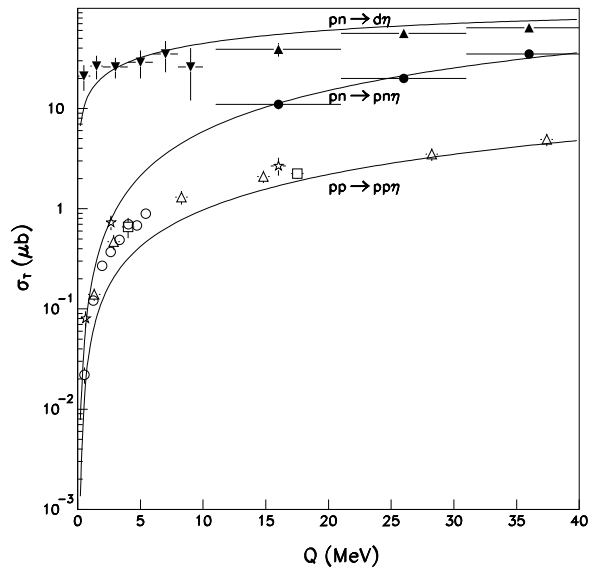}}
\caption{Variation of the total cross sections for the $pp\to pp\eta$,
$pn\to pn\eta$, and $pn\to d\eta$ reactions with excess energy. The
data are taken from references \protect\cite{Pinot} (square),
\protect\cite{Calen1} (open triangle), \protect\cite{Hibou} (star),
\protect\cite{Cosy11} (open circle), \protect\cite{Calen5} (closed circle),
\protect\cite{Calen2} (closed triangle), and \protect\cite{Calen3}
(inverted triangle). The curves are the model predictions with 
standard parameters, evaluated without $\eta$ exchange.}
\end{center}
\label{fig5}
\end{figure}

It was suggested several years ago~\cite{GW} that the cross section
ratio could be increased by invoking a destructive
interference between $\rho$ and $\pi$ exchange in the $I=1$ channel, and
hence constructive in the $I=0$. With the values of the parameters given
in Section~4, we find the $\rho$-amplitude to be about three times stronger
than that of the $\pi$, with the $\omega$ term being another factor of three
smaller. These values then lead to the broad overall agreement with the
$pp\to pp\eta$ and $pn\to pn\eta$ total cross sections shown in
Fig.~5. The $pn\!:\!pp$ ratio is a little too large 
but, if the $\rho$ coupling were increased by a mere 5\%, then this
would be reduced to the experimental factor of around $6.5$~\cite{Calen5}.

Whereas the absolute values of the cross sections and the $pn\!:\!pp$
ratio depend very sensitively upon the parameters, the same is not true
for the shape of the energy dependence of the $pp\to pp\eta$
total cross section where the near-threshold values need to be enhanced.
The $\eta$-nucleon final state interaction, that we
have neglected in our analysis, could well supply such an
enhancement. The incorporation of such an effect is important but not
straightforward and we have made no attempts in this direction.

\vspace{-3mm}
\noindent
\input epsf
\begin{figure}[ht]
\begin{center}
\mbox{\epsfxsize=3.5in \epsfbox{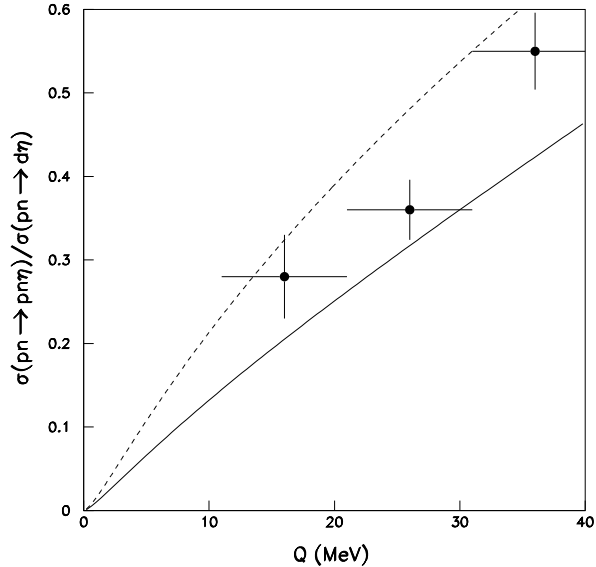}}
\caption{Ratio of the $pn\to pn\eta$ and $pn\to d\eta$ total cross
sections at the same excess $Q$. Data from
Ref.~\protect\cite{Calen5,Stina,Calen2} are compared to the full
calculation (solid curve) and the simplified final-state-interaction
approach of Eq.~(\protect\ref{6_6}) (broken curve) which only includes
$I=0$ contributions.}
\end{center}
\label{fig6}
\end{figure}

The largest cross section shown in Fig.~5 near threshold is that for 
$pn \to d\eta$, but its dominance is mainly a reflection of the faster
rise of the two-body phase space as compared to those of the three-body
$pp\eta$ and $pn\eta$ final states. To see this effect in greater
detail, we show in Fig.~6 the ratio of the $pn\to pn\eta$ to the 
$pn\to d\eta$ total cross section on a linear scale, where the effect of
the rapid
phase-space variation can be clearly seen. The theoretical prediction
of Eq.~(\ref{6_6}), based upon the extrapolation theorem~\cite{FW3},
overestimates the ratio and this would be made slightly worse if
the small $I=1$ contribution were also included. The situation is changed
in the full calculation and this is mainly due to the deuteron $D$-state
which increases the $np\to d\eta$ estimate by up to a factor of two.
This large effect is due to the deuteron wave function occuring linearly
in Eq.~(\ref{6_3}).

Any residual discrepancies in Fig.~6 may be partially experimental in
origin since the data were obtained using a deuterium target and the
values of $Q$ deduced by kinematic fitting~\cite{Calen5,Stina,Calen2}.
The uncertainties in the $pn\eta$ final state are particularly large and
one cannot rule out a systematic shift of a few MeV. In order to get
better resolution in $Q$, it would be an advantage to measure the
spectator proton directly. Such a procedure is now feasible~\cite{Tord}.

The results shown in Fig.~5 have been obtained after neglecting
$\eta$ exchange. Now, because of the large $\eta N$ elastic amplitude,
pure $\eta$ exchange with an $\eta NN$ coupling of $G_{\eta}^{\,2}/4\pi = 1$
would lead to a $pp\to pp\eta$ cross section twice as big as that
for $\pi$ exchange. If the other parameters were not altered, a value of
$G_{\eta}^{\,2}/4\pi = 0.25$ would reduce the $pn\!:\!pp$ ratio prediction
to below 5. To reproduce this experimental charge ratio with
$G_{\eta}^{\,2}/4\pi = 1$, the $\rho$-exchange amplitude would have to be 
reduced by a factor of two, and vector meson dominance normally leads
to estimates that are far more reliable than that. Since the charge
ratio is only weakly affected by initial distortion or choice of $NN$ 
wave functions or form factor parameters \textit{etc.}, our model 
indicates that $G_{\eta}^{\,2}/4\pi < 1$ and is probably much smaller. 

The $pp\to pp\eta$ cross section is a function of
$\cos^2\theta_{\eta}$ because the two initial protons are identical.
From Eq.~(\ref{3_3_5})
and (\ref{4_1_2}), it is seen that to first order in the $d$-wave
amplitudes the differential cross section should be linear in
$\cos^2\theta_{\eta}$. This is illustrated in Fig.~7, where the
predictions, without $\eta$ exchange, have been increased by
25\% before comparing them with the results of~\cite{Calen4}. 

The agreement with experiment is reasonable, except for the largest angle
point. The large slope
parameter quoted in~\cite{Calen4} is due to the weight of this point
in the free fit to the data and we see no way whereby such a
value could be achieved within the current approach. It should also be
noted that this point affects the overall scale, because the angular
distribution is normalised to the integrated cross section found in previous
studies~\cite{Calen1}.

\noindent
\input epsf
\begin{figure}[ht]
\begin{center}
\mbox{\epsfxsize=3.5in \epsfbox{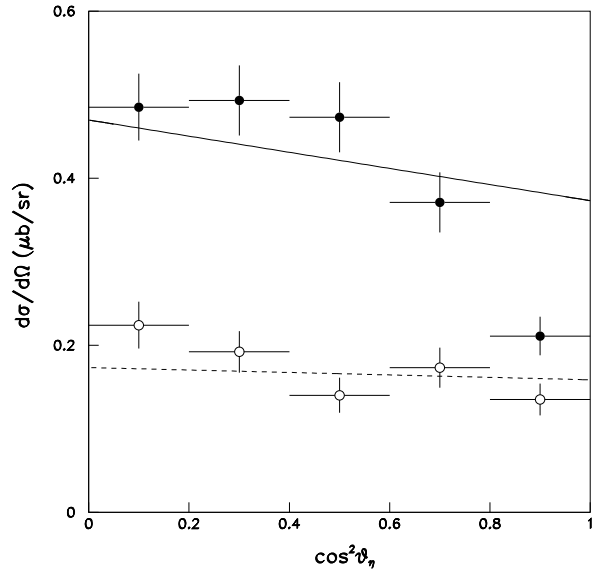}}
\caption{Distribution in the c.m.\ angle of the $\eta$ in the
$pp\to pp\eta$ reaction at $Q=37$~MeV (closed circles) and
$Q=16$~MeV (open circles) from~\protect\cite{Calen4}. The predictions of
our model with standard parameters without $\eta$ exchange, shown as the
corresponding solid and broken lines respectively, have been increased by a
factor of $1.25$.}
\end{center}
\label{fig7}
\end{figure}

Though small, the slope is still much larger than that found for
$\gamma p\to \eta\,p$~\cite{Krusche} or $\pi^-p\to \eta\,n$~\cite{Ben3}.
It is here important to note that the photoproduction data are maximal
for $\theta_{\eta}\approx 90^{\circ}$ where the pion-production results
show a minimum. To get agreement with the $pn/pp$ total cross section
ratio, the $\rho$ and $\pi$ contributions must cancel in the $pp$ case,
and this enhances the curvature and leads to the agreement shown in
Fig.~7. For $pn\to pn\eta/d\eta$ the interference is
constructive and we would expect the slope to be an order of magnitude
smaller.

There are, as yet, no data on the analysing power $A_y$ in $pp\to pp\eta$,
though there is an approved proposal at COSY-11 to attempt the first
investigation~\cite{Walter}. There are also no measurements of $A_y$ in
$\pi^-p\to \eta\,n$ but, if we consider only $\rho$ exchange, we predict
that $A_y$ for $pp\to pp\eta$ should be of the same size as that
for $\gamma\,p\to\eta\,p$~\cite{Bock} at the same value of the $\eta$-$N$
momentum. With the parameterisation of Eqs.~(\ref{3_4_11}, \ref{3_4_12}),
one expects that the analysing power to vary as
\begin{equation}
\label{7_2_1}
A_y = 2\,A_y^{\textrm{\scriptsize max}}\,
\sin\theta_{\eta}\,\cos\theta_\eta\:.
\end{equation}

\vspace{-3mm}
\noindent
\input epsf
\begin{figure}[ht]
\begin{center}
\mbox{\epsfxsize=3.5in \epsfbox{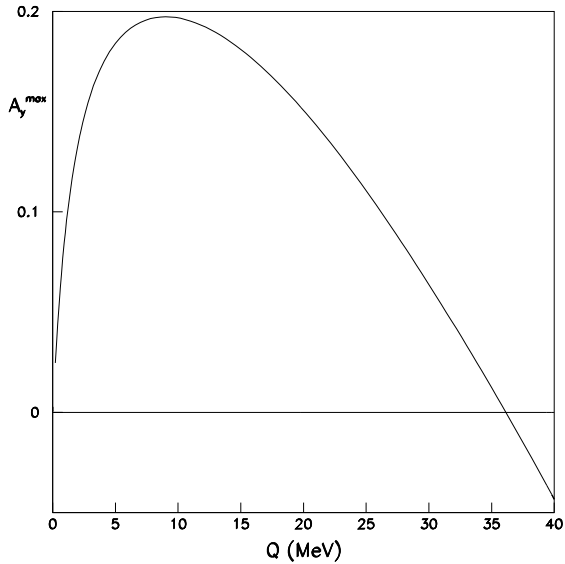}}
\caption{Predicted value of the $pp\to pp\eta$ maximum analysing power
coefficient of Eq.~(\protect\ref{7_2_1}) as a function of excess energy,
estimated assuming only vector-meson exchange.}
\end{center}
\label{fig8}
\end{figure}

The predicted energy dependence of the maximum value of the analysing power
$A_y^{\textrm{\scriptsize max}}$ is shown in Fig.~8. The values are small
because of the node in the input of Eq.~(\ref{3_4_12}) at $\eta\approx 0.4$,
and so they may be sensitive to the neglected $p$-wave and pion-exchange
terms. It may be of interest to note, however, that the maximum occurs for
$Q\approx 10$~MeV, which is a region where the acceptance of the COSY-11
spectrometer is significant~\cite{Walter}.
\newpage
\section{Conclusions}
We have been able to describe, within a one-meson-exchange picture, 
the main features of the experimental data on $\eta$ production in
nucleon-nucleon collisions near threshold.
Although most of the energy dependence is reproduced,
this is mainly an effect of phase space, combined with the
nucleon-nucleon final-state interaction. The remaining low-$Q$ deviations
in both the $NN\to NN\eta$ and $pn\to d\eta$ reactions could be due
to the $\eta$-nucleon final-state interaction that was not included in the
calculation.

The good agreement with the magnitude of the $pp\to pp\eta$ total cross
section may, in part, be fortuitous given the uncertainties in the
the coupling constants and other parameters, including those arising
from the use of vector-meson dominance to extract the input amplitudes. 
Although we have not adjusted the input in order to fit the data, it 
should be stressed that there remains significant flexibility in the 
numerical predictions arising from these uncertainties. 

A more serious worry is the sensitivity to the short-range variation of
the nucleon-nucleon wave functions arising from the form of the production
operators. The Paris $NN$ potential~\cite{Paris}, that we have used in all
our estimates, is essentially local and this has a significant repulsive
core that suppresses the wave functions at short distances. On the
other hand, the Bonn potential~\cite{Bonn,Machleidt} has a non-local
component that allows the core to be much softer and this results in much
less small-$r$ suppression. This means that the evaluation of the
$pn\to d\eta$ cross section on the basis of the integrals in
Eqs.~(\ref{6_2}, \ref{6_3}) is about three times bigger for the Bonn 
deuteron wave function than that of Paris. The difference would be even
larger without form factors to regularise the production operators at short
distances. We would argue that, because we are using local production
operators, then it is more logical to use wave functions corresponding
to a local potential. The $np\to d\eta$ cross section is sensitive to
the deuteron $D$-state wave function at short distances and the ratio
of this to the $S$-state also differs significantly between Paris and Bonn.

In our approach, the $\rho$-exchange amplitude is bigger than
that of the $\pi$ and it is particularly reassuring that the
destructive interference required to give good agreement for the ratio
of the $pn\to pn\eta/pp\to pn\eta$ cross sections is such as to
give the correct shape of angular dependence in the $pp\to pp\eta$
case. It should be noted that if the $\pi$-exchange term were to dominate,
then the slope of the differential would have the sign opposite to that
of experiment. This illustrates the importance of analysing all the
$\eta$-production channels simultaneously.

The uncertainty in the $\eta NN$ coupling constant
makes it impossible to estimate the $\eta$-exchange term with any
reliability. It cannot be too large without destroying the good
agreement with the $pn/pp$ ratio and this allows us to derive the upper
limit \textit{within the model} of $G_{\eta NN}^2/4\pi < 1$. It should 
be noted in this context that the $\eta$ contribution in 
semi-phenomenological nucleon-nucleon potentials~\cite{Bonn,Paris} is
to be associated with the exchange quantum numbers and not necessarily 
purely with the true $\eta$-meson.

Although our predictions for the slope in $pp\to pp\eta$ are
compatible with the data, it must be admitted that the experimental
measurements are still far from definitive in both statistical and
systematic aspects. A significant improvement is to be expected in
the results through the use of the WASA detector~\cite{WASA}. This is
designed for the study of rare decays of the $\eta$ but the production
of the meson will also be investigated as a by-product under very good
conditions. One test of our model is the prediction that the $\eta$
angular distributions in the proton-neutron cases should be much
flatter than for proton-proton. This could be checked in the
$pn\to d\eta$ reaction, where the measurement of all fast final-state
particles means that the kinematics may be well defined even without
detecting the spectator proton~\cite{WASA}. The only measurements extant
showed an $\eta$ distribution that was consistent with
isotropy~\cite{Stina}.

Another partial check on our model could come from a measurement of
the beam analysing power in $pp\to pp\eta$. This is predicted to
have a maximum, albeit rather small, at quite low values of $Q$. The
predictions are not conclusive because of the lack of $\pi^-p\to\eta\,n$
analysing power input to complement the extensive unpolarised
differential cross section data which should appear in the near
future~\cite{Ben1}.

The gravest drawback in the calculation is the neglect of $\eta$
rescattering. There are as yet no well-behaved $\eta$-nucleon
potentials in the literature and the short range of the
$\eta$-production operator means that this effect will depend upon
the details of the $\eta$-nucleon interaction and not just the
on-shell $t$-matrix. A consistent three-body treatment of the
$\eta\,NN$ final state in $\eta$ production is a challenge still
to be met.\\

We are grateful to members of experimental groups at Saclay, CELSIUS,
and COSY for discussions about $\eta$ production over many years. The
present work has been helped by correspondence with T.~Pe\~{n}a
and L.~Tiator. S.~Prakhov and B.M.K.~Nefkens provided us with valuable
information on the Crystal Ball programme. This work has been supported by
the Royal Swedish Academy of Sciences, within the framework of the European
Science Exchange Programme. One of the authors (CW) would like to thank the
The Svedberg Laboratory for its generous hospitality.
We would like to dedicate the paper to our friend Nimai Mukhopadhyay 
with whom we shared a common interest in $\eta$-production over many years.

\end{document}